\documentclass[12pt]{article}

\usepackage[utf8]{inputenc}
\usepackage{charter}
\usepackage{fullpage}
\usepackage{hyperref}
\usepackage{graphicx}
\usepackage{amssymb,amsmath}
\usepackage{lipsum}
\usepackage[sort&compress,numbers]{natbib}
\usepackage{hyperref}
\hypersetup{hidelinks}
\usepackage{wrapfig}
\usepackage[usenames,dvipsnames,table]{xcolor}
\usepackage[total={6.5in,9in},top=1in,headsep=0.1in,headheight=1in]{geometry}
\usepackage{lineno}
\usepackage{geometry}
\usepackage{fullpage}
\usepackage[T1]{fontenc}    
\usepackage{hyperref} 
\usepackage{bm}	
\usepackage[normalem]{ulem}
\usepackage{caption}
\usepackage{subcaption}
\usepackage{verbatim}
\usepackage{lineno}
\usepackage{url}    
\usepackage{booktabs}       
\usepackage{amsfonts}       
\usepackage{nicefrac}
\usepackage{natbib}
\usepackage{xcolor}
\usepackage{aas_macros}

\newcommand\snowmass{
\begin{center}
  \rule[-0.2in]{\hsize}{0.01in}\\
  \rule{\hsize}{0.01in}\\
  \vskip 0.1in
  Submitted to the Proceedings of the US Community Study\\ 
  on the Future of Particle Physics (Snowmass 2021)\\
  \rule{\hsize}{0.01in}\\
  \rule[+0.2in]{\hsize}{0.01in}\\[-2em]
\end{center}
}
 
 \usepackage[firstpage=true]{background}
\backgroundsetup{contents={\parbox{6.5in}{\snowmass}}, scale=1,placement=top,opacity=1,color=black,position={3.25in,1.2in}}

\usepackage{fancyhdr}
\fancypagestyle{plain}{%
  \fancyhf{}%
  \fancyhead[C]{}
  \fancyfoot[C]{\thepage}
}

\fancypagestyle{empty}{%
  \fancyhf{}%
  \fancyhead[C]{{\it Snowmass White Paper: Machine Learning and Cosmology}}
  \fancyfoot[C]{\thepage}
}

\usepackage{authblk}

\author[1]{Cora Dvorkin\footnote{cdvorkin@g.harvard.edu}}
\affil[1]{Department of Physics, Harvard University, 17 Oxford Street, Cambridge, MA 02138, USA}
\author[1,2,3]{Siddharth Mishra-Sharma\footnote{smsharma@mit.edu}}
\affil[2]{The NSF AI Institute for Artificial Intelligence and Fundamental Interactions}
\affil[3]{Center for Theoretical Physics, Massachusetts Institute of Technology, Cambridge, MA 02139, USA}
\author[4,5,6]{Brian Nord\footnote{nord@fnal.gov}}
\affil[4]{Fermi National Accelerator Laboratory, Batavia, IL 60510, USA}
\affil[5]{Department of Astronomy and Astrophysics, University of Chicago, IL 60637, USA}
\affil[6]{Kavli Institute for Cosmological Physics, University of Chicago, Chicago, IL
60637, USA}
\author[7,8,9]{V. Ashley Villar\footnote{vav5084@psu.edu}}
\affil[7]{Department of Astronomy \& Astrophysics, The Pennsylvania State University, University Park, PA 16802, USA}
\affil[8]{Institute for Computational \& Data Sciences, The Pennsylvania State University, University Park, PA, USA}
\affil[9]{Institute for Gravitation and the Cosmos, The Pennsylvania State University, University Park, PA 16802, USA
}
\author[10]{Camille Avestruz}
\affil[10]{Department of Physics; University of Michigan, Ann Arbor, MI 48109, USA}
\author[11]{Keith Bechtol}
\affil[11]{Physics Department, 2320 Chamberlin Hall, University of Wisconsin-Madison, 1150 University Avenue Madison, WI 53706-1390}
\author[4]{Aleksandra \'Ciprijanovi\'c}
\author[12]{Andrew J. Connolly}
\affil[12]{Department of Astronomy, University of Washington, Seattle, WA, 98195, USA}
\author[13]{Lehman H. Garrison}
\affil[13]{Center for Computational Astrophysics, Flatiron Institute, Simons Foundation, 162 Fifth Ave., New York, NY 10010, USA}
\author[14]{Gautham Narayan}
\affil[14]{University of Illinois at Urbana-Champaign Urbana, IL, USA}
\author[15,16]{Francisco Villaescusa-Navarro}
\affil[15]{Center for Computational Astrophysics, Flatiron Institute, 162 5th Avenue, New York, NY, 10010, USA}
\affil[16]{Department of Astrophysical Sciences, Princeton University, Peyton Hall, Princeton NJ 08544, USA}

\date{}
\title{Machine Learning and Cosmology}

\begin{document}

\maketitle

\begin{abstract}
Methods based on machine learning have recently made substantial inroads in many corners of cosmology. Through this process, new computational tools, new perspectives on data collection, model development, analysis, and discovery, as well as new communities and educational pathways have emerged. Despite rapid progress, substantial potential at the intersection of cosmology and machine learning remains untapped. In this white paper, we summarize current and ongoing developments relating to the application of machine learning within cosmology and provide a set of recommendations aimed at maximizing the scientific impact of these burgeoning tools over the coming decade through both technical development as well as the fostering of emerging communities.
\end{abstract}

\newpage 

\section{Introduction}

The interplay between models and observations is a cornerstone of the scientific method, aiming to inform which theoretical models are reflected in the observed data. Within cosmology, as both models and observations have substantially increased in complexity over time, the tools needed to enable a rigorous comparison have required updating as well. With an eye towards the next decade in cosmology, the vast data volumes to be delivered by ongoing and upcoming surveys, as well as the ever-expanding theoretical search-space, motivate a re-thinking of the statistical machinery used. In particular, we are now at a crucial juncture where we may be limited by the statistical and data-driven tools themselves rather than the quality or volume of the available data. 

Methods based on artificial intelligence (AI) and machine learning (ML) have recently emerged as promising tools for cosmological applications, demonstrating the ability to overcome some of the computational bottlenecks associated with traditional statistical techniques. Machine learning is starting to see increased adoption across different sub-fields of and for various applications within cosmology. At the same time, the nascent and emergent nature of practical artificial intelligence motivates careful continued development and significant care when it comes to their application in the sciences, as well as cognizance of their potential for broader societal impact. 

In this white paper, we provide an overview of some of the ways machine learning methods are becoming increasingly central to the way cosmological data is collected, analyzed, and interpreted. Along the way, we highlight our vision for necessary developments, framing these as recommendations---both technological as well as sociological---for the widespread safe and equitable adoption of machine learning methods within cosmology in the coming decade.

\section{Examples of science cases}

\subsection{Cosmic Probes}

Cosmology is the study of the content and evolution of our Universe.  In the $\Lambda$CDM standard model of cosmology, the Universe has three primary components: dark energy (the source of the accelerated expansion of our Universe), dark matter (comprising the majority of the mass density in our Universe, whose presence thus far has been inferred through gravitational interactions), and ordinary visible matter.
While the observational data has thus far remained largely consistent with the standard model of cosmology, fundamental physics questions remain unanswered.  Unknowns include the particle nature of dark matter, the source of the accelerated expansion of the Universe, and the physics that seeded the first structures in the Universe. 

There is an exciting prospect for many of the outstanding problems in cosmology to be solved, or substantial progress towards their solution to be made, using the statistical power offered by current and upcoming cosmological experiments.   
Maximizing the scientific return of forthcoming data will require methods that can extract as much information as possible from observations while controlling for systematics associated with measurements as well as theoretical models.

In multiple subfields of cosmology, the pathway to cosmological constraints generally encompasses a source identification step, a summary measurement, and a comparison of the observation to theoretical models for parameter inference.
In the rest of this section, we outline examples of cosmic probes where machine learning has already made a significant impact, with continued progress expected over the next decade as observations continue to increase in sensitivity. We note that the directions and works highlighted in this as well as subsequent sections are meant to be representative, rather than exhaustive.

Machine learning techniques will be crucial for detecting and classify cosmological sources, extract information from images, and optimize observing strategies. Examples of probes of cosmology include galaxy clustering, supernovae, strong and weak gravitational lensing, and the cosmic microwave background.        
Cosmology analyses that incorporate multiple probes have increased constraining power for two primary reasons. First, different probes span model space parameters in complementary ways, allowing for tighter constraints of cosmological parameters.  Second, each observable is impacted by its own set of astrophysical processes and observational systematics. Machine learning methods can be used to fully realize the potential of multi-probe cosmology by decorrelating the effect of systematics and optimally combining information from multiple surveys~\cite{2021arXiv210706984J}.  

Many scientific analyses that aim to extract cosmological information rely on astronomical catalogs.  These catalogs are constructed from astronomical images, with their contents usually reflecting a ``best-fit" model of the underlying image constituents (rather than a distribution over possible constituents).  The associated information loss folds in information about uncertainty that cannot be retained in a downstream analysis. Machine learning has the potential to fully realize probabilistic cataloging~\cite{2013AJ....146....7B,Daylan:2016tia}    
which aims to detect and characterize constituents (e.g., stars and galaxies) from astronomical images in a fully probabilistic manner. These capabilities were recently demonstrated in the context of deblending crowded stellar fields~\cite{2021arXiv210202409L} and can significantly enhanced the scientific output of cosmological surveys in the forthcoming decade.

The detection and characterization of strong gravitational lenses has emerged as a promising application of machine learning methods. An early application included a community-driven challenge on detecting strong lenses (see Ref.~\cite{Metcalf:2018elz} for a summary), in which machine learning methods were shown to outperform traditional statistical techniques. Detection algorithms for strong lenses were subsequently applied to data from the Dark Energy Survey (DES)~\cite{2019MNRAS.484.5330J,DES:2021zcj} and the Dark Energy Spectroscopic Instrument (DESI)~\cite{2021ApJ...909...27H,Stein:2021vdx}. Further work~\cite{Hezaveh:2017sht, PerreaultLevasseur:2017ltk, Morningstar:2018ase} proposed applying parameter inference and uncertainty quantification methods in order to characterize the properties of lensed sources and lensing galaxies. More recently, with an eye towards the large sample of gravitational lenses that will be imaged by forthcoming cosmological surveys like \emph{Euclid} and LSST, there has been significant effort towards understanding how to utilize machine learning to optimally exploit this data towards source/lens characterization~\cite{Wagner-Carena:2020yun,Chianese:2019ifk,Karchev:2021fro,Morningstar:2019szx}, Hubble constant inference~\cite{Park:2020eat}, and characterization of dark matter substructure within the lensing galaxies~\cite{Daylan:2017kfh,Alexander:2019puy, DiazRivero:2019hxf, Alexander:2020mbx, Ostdiek:2020mvo, Ostdiek:2020cqz, Coogan:2020yux, He:2020rkj, Legin:2021zup, Alexander:2021gxq, Wagner-Carena:2022mrn,Brehmer:2019jyt} in a scalable manner.

Weak gravitational lensing is another area where machine learning has shown significant advantages over traditional methods. In particular, machine learning methods allow for cosmological parameter inference via weak lensing measurements by leverage information beyond conventional one- and two-point statistics~\cite{Jeffrey:2020xve, Gupta:2018eev,Fluri:2018hoy,Ribli:2019wtw,Fluri:2019qtp,Mootoovaloo:2020ott,Lu:2021eeh}. Additionally, machine learning methods have demonstrated the ability to outperform traditional estimators in mass mapping---reconstructing the weak lensing convergence maps from observed galaxy ellipticities~\cite{Tewes:2018she,Springer:2018aak,Jeffrey:2019fag,Remy:2022ixn}.

Although the primary Cosmic Microwave Background (CMB) signal in the standard cosmological scenario can be statistically described as a Gaussian random field and efficiently analyzed with angular power spectrum estimators, machine learning methods have demonstrated superior performance on CMB analysis tasks including CMB lensing reconstruction~\cite{Caldeira:2018ojb}, foreground separation~\cite{2021MNRAS.500.3889A,Petroff:2020fbf} and inference with polarization maps~\cite{Jeffrey:2021fcg,Guzman:2021ygf}. Given the anticipated complexity of data, these techniques can significantly enhance the deliverable output of forthcoming surveys like Simons Observatory~\cite{SimonsObservatory:2018koc}, CMB-S4~\cite{CMB-S4:2016ple}, and proposed high-resolution experiments like CMB-HD~\cite{Sehgal:2019ewc}. Analysis of CMB maps using machine learning techniques has also been explored for characterizing galaxy clusters via Sunyaev–Zel'dovich and CMB lensing signatures~\cite{Lin:2021dim,deAndres:2021tjl,Gupta:2020yvd,Gupta:2020him}.

Machine learning has also been proposed as a way to confront the complexity of 21-cm neutral hydrogen data probing the epoch of reionization, with applications explored for parameter estimation~\cite{Wadekar:2020oov,Hassan:2019cal,Neutsch:2022hmv,Sikder:2022hzk,Hortua:2020ljv} and signal extraction~\cite{Makinen:2020gvh,Villanueva-Domingo:2020wpt,Ghosh:2020fpy}.

\subsection{Time Domain \& Multi-messenger Astrophysics}

In the time domain, machine learning techniques have a key role to play in the era of wide-field surveys across the electromagnetic and gravitational wave spectrum. Of these surveys, the {\it Vera C. Rubin Observatory}'s Legacy Survey of Space and Time (LSST) will produce $\sim$10 million alerts from time-domain phenomena every night---higher than can conceivably be inspected visually by the entire astronomical time-domain community in a lifetime. LSST will be joined by the next-generation Very Large Array (ngVLA), CMB-S4, the LIGO-VIRGO-Kagra Collaboration and numerous other data-intensive experiments over the next decade. Each of these will provide us a different, multi-messenger window into the variable sky and the cosmos.

The rate of transients alerts from these wide-field surveys far surpasses available spectroscopic resources \emph{already}. Less than 5\% of transient events reported to the International Astronomical Union's Transient Name Server (TNS) are followed up spectroscopically. The order-of-magnitude increase in alerts with LSST will dramatically increase the stress on available spectroscopic resources. In a cosmological context, Type Ia supernovae are especially important for measurements of the Hubble constant at a unique rung of the distance ladder. ML techniques will be essential to classify these events in real time (for active, spectroscopic follow-up) or for archival analysis. 

While there has been significant progress in developing ML methods to characterize and classify events in real-time from wide-field survey data (e.g. alert broker systems such as NOIRLab's ANTARES~\citep{ANTARES:2018uvq, 2021AJ....161..107M}, as well as algorithms such as RAPID~\citep{Muthukrishna:2019wgc}, SuperNNova~\citep{2020MNRAS.491.4277M}, SuperRAENN~\citep{Villar:2020epn}), much is left to be done before LSST in order to optimally understand the \textit{diversity} of events we expect to observe. We note that the ML community has also developed many more sophisticated techniques (e.g., transformers, latent stochastic differential equations, neural processes) which will greatly benefit our domain challenges in the near future.

There is also a major role for unsupervised learning methods in the coming decade, to identify rare events that have never been seen before (anomaly detection), and are buried within the alert stream \cite{li2021preparing}. Indeed, while only one kilonova event---GW170817---has been identified in both gravitational waves and the electromagnetic spectrum, we should expect LSST to discover many kilonovae that are beyond LIGO-Virgo-Kagra's detection limit \cite{andreoni2021target}. These unsupervised learning methods can also be used to identify events that were missed during real-time processing, and the promise of these methods has already been demonstrated by teams processing Pan-STARRS, Dark Energy Survey (DES) and Zwicky Transient Facility (ZTF) data (e.g. \citep{2021MNRAS.502.5147M}).

Finally, the key promise of wide-field surveys is understanding \emph{populations} of objects in the time-domain. Modeling these populations to infer fundamental physics (e.g. inferring cosmological parameters from type Ia supernovae) is a complex multi-level modeling (sometimes known as hierarchical Bayesian) problem, and these analyses have either typically made simplifying assumptions in their models to make them tractable to evaluate with traditional inference techniques such as Markov Chain Monte Carlo sampling. However, these simplifying assumptions themselves will lead to systematic biases. Building blocks of machine learning models, such as deep neural networks, can be effectively leveraged within schemes like variational inference in order to perform approximate inference on models defined in high-dimensional spaces. We expect that analysis groups will need to complement traditional techniques on high-performance computing resources with new methods that make much greater use of GPUs.

Multi-messenger Astrophysics (MMA) takes all of the requirements we have detailed above and adds additional complexity. Sources that are energetic enough to distort the fabric of space-time and cause gravitational waves are intrinsically rare, and a large subset of those events that also emit electromagnetic radiation (nearby core-collapse supernovae, neutron star mergers, and the merges of neutron stars and black holes) also evolve extremely quickly. For instance, the canonical MMA event, GW170817, will only have 1--2 detections in a survey such as LSST with its currently planned cadence \cite{cowperthwaite2019lsst}. 

Detecting and characterizing MMA events therefore requires discovery of very rare events with very sparse and heterogeneous data from multiple facilities all in real-time. ML methods, including unsupervised anomaly detection techniques \cite{morawski2021anomaly}, and hybrid architectures such as convolutional recurrent neural networks (CRNNs) \cite{gebhard2019convolutional, krastev2021detection}, and simulation-based inference techniques \cite{dax2021real} will be necessary to process the smorgasbord of observations from different facilities, flag these events within hours, and automatically trigger follow-up studies. Beyond the ML techniques highlighted in this white paper, MMA will require significant investment in cross-survey cyberinfrastructure to help the community store, process and share a mix of public and private data in order to understand these enigmatic events.   

\section{Computing and Data}

\subsection{Data Processing}

During and after data acquisition of modern, large-scale cosmological surveys, there are multiple opportunities to improve the speed and accuracy of data processing. 
We will approach these items sequentially in the nominal processing steps.

First, we envision new modalities for smart triggering---or fast selection and analysis of data of critical elements upon observation.
This is important especially for the detection of transient or time-varying data that requires timely follow-up. 
Currently, this is typically handled highly successfully with difference-imaging (e.g., Antares in LSST). 
However, the predictive capabilities of this method are limited in terms of the accuracy and prediction of early detection, which constrains our understanding of which types of objects are being selected. 
Being able to predict the class of objects with fewer observations greatly increases the probability of capturing objects before they fade away.
In general, earlier understanding of object status and classification will allow us to choose the most valuable objects for further observation. 
For objects that are not transient, but uniquely important for other science cases, fast and accurate identification will permit for greater coordination with other observational resources. 
Future solutions include embedding deep learning algorithms within specialized computational units like TPUs and FPGAs, which are designed for hyperfast data processing. 
This is the model under current usage within particle physics programs at the Large Hadron Collider. 
Preliminary work has taken place in astronomy to demonstrate this capability~\cite{2020AAS...23530303C}.
We envision a possible future where these fast processors are used to pre-process raw data for fast detection: they may be part of an array at the facility where the long-term data processing is occurring, or they may be located at the observational facility.
In the latter case, this would bypass data transfer times, as well as facilitate adaptive action for immediate updates of the observing schedule from observation to observation.

Next, consider the cleaning and calibration of data.
Multiple types of noise artifacts and confounding patterns can degrade image quality: sensor artifacts, particles (e.g., cosmic rays), bright foreground stars, point spread function variation, galaxy blending, etc. 
All of these need to be modeled and removed from the data before science can be performed.
The first obvious usage for machine learning is the removal of basic artifacts like cosmic rays, which has already been demonstrated~\cite{2020ApJ...889...24Z}. 
Foreground stars can be detected with tools like Source Extractor, which uses machine learning for e.g., source classification~\cite{1996A&AS..117..393B}. 
Patterns in sensors (like the tree rings or brighter-fatter pixels discovered in the Dark Energy Survey data) are more difficult to discover and typically require a great deal of data investigation: replacing this task with algorithmic tools that can discover these pattern has the potential to save a great deal of commissioning and science verification time. 
Potential tools for this include detailed sensor simulators that can generate patterns in the sensors as propagated to data, as well as inference loops that can test many flexible data models and propose sensor patterns as systematics.
Object blending is already a significant challenge for surveys like LSST; the fainter and deeper our imaging becomes, the worse this problem will be. 
This is currently a feature case study and much progress has been made~\cite{2021arXiv210202409L}, and we encourage continued work on this challenge.

\subsection{High-Performance Computing}
\label{subsec:HPC}

We anticipate high-performance computing (HPC) to play a crucial role in our capability to use ML for cosmology this decade. The massive amount of data from cosmological surveys such as the Dark Energy Spectroscopic Instrument (DESI), \textit{eROSITA}, \textit{Euclid}, \textit{Roman Space Telescope}, the {\it Vera C. Rubin Observatory}, Simons Observatory, and the Square Kilometre Array (SKA) will require dedicated infrastructure to host the data.  Likewise, the complexity of the models used to interpret these surveys demands significant computational and storage capabilities.

Consider the example of training an ML model to interpret a galaxy survey by comparison to cosmological simulations. Every stage of this process is an HPC-intensive task: (1) simulations must be produced at large HPC facilities, using $\mathcal{O}$(100K-1M) GPU-hours (e.g.~AbacusSummit \citep{2021MNRAS.508.4017M}; Uchuu \citep{2021MNRAS.506.4210I});  or $\mathcal{O}$(10M-100M) CPU-hours \citep{Quijote, IllustrisTNG, ASTRID, CAMELS} (2) outputs are stored on large, high-throughput file systems, totalling $\mathcal{O}$(1 PB); (3) the analysis (generation of data products) is executed in cluster environments (often using CPUs) requiring a small percent of the cost incurred in running the simulations; (4) ML models are trained on these data products (using GPUs or other accelerators).  The next decade will only see these requirements increase, as upcoming surveys explore both broader sky area and greater depth, demanding simulations with both larger volume and finer resolution. This will increase the HPC expense of every step of this process.

However, ML does not just place demands on HPC, but also offers opportunities.  Traditional simulations, or parts of them, can be accelerated with ML \citep{He_2019, deoliveira2020fast, Kaushal_2021}.  High-resolution simulations may be emulated based on low-resolution ones \citep{Doogesh_2020, Yin_2020, Yueying_2021}; expensive physical calculations replaced with a machine learning interpolation thereof \citep{CAMELS}. This may increase the slope of the relation between computational resources and the size and resolution of the simulations.  However, this will pose significant challenges to the storage and manipulation of this data, that we expect to be of the order of tens to hundreds of petabytes.

In order to maximize the usage and utilization of simulations and observations, large collections of data must be placed in a publicly accessible location in national or international HPC centers where users can access and work directly with the data without creating copies of it elsewhere. At the same time, subsets of this data should be easy to download, as interesting investigations can often be done on just a few MB of data products.

Cloud computing may be a compelling option for some HPC applications, with its massive storage and distributed compute resources. Cloud computing also offers an opportunity to present users with a standard software stack, promising a solution to the byzantine task of setting up an ML-ready software environment.  However, applications like simulations or some kinds of ML training that need efficient communication between many worker nodes may not run well in cloud environments. The opportunities for HPC-in-cloud will surely expand in the next decade, though, driven by ML applications, and efforts to leverage it for scientific workflows should be supported.

Ensuring reproducible research in an HPC context will require new ways for users to interact with HPC resources. Reproducibility efforts in astrophysics are now widespread, and span a spectrum from adopting software-industry testing and continuous integration practices, to fully reproducible workflows that allow readers to reproduce a paper's results (e.g.~generate all values and figures) at the click of a button 
These pipelines tend to be cloud-hosted and use small datasets (few GB) and only a few CPU-hours---far below ML requirements. But the transparency and pedagogical value of such workflows cannot be ignored by the cosmology ML community. HPC centers must offer entry points to their resources that allow automation of these workflows, including access to large datasets and compute resources (especially GPUs).

The increasing demanding computational needs in astrophysics may trigger synergistic collaborations between scientists and the industry. For instance, the methods used in the industry to compress certain classes of data (e.g. videos in YouTube) will reduce the storage needs in astrophysics, enabling deeper and more complete analyses of the data.

In summary, in the next decade HPC must support multiple workflows to maximize science at the intersection of ML and cosmology.  The computational infrastructure must enable flagship efforts, such as hosting large observational data sets, running massive cosmological simulations, and training ML models to map between the two.  HPC centers must also make such data (simulation and observations) publicly available, such that users can bring their computations to the data or download narrow subsets of the data to support local workflows.  The software side will require dedicated, expert effort in the areas of cloud hosting, containerization, and software environments to bridge the gap between the code scientists write and the potential performance HPC hardware promises.

\section{Simulations}

While humans can easily identify patterns in low-dimensional data, machine learning algorithms can also perform this task in high-dimensional spaces. Thus, simulations can be used as a laboratory to identify unrecognized patterns that can help our understanding of the underlying mechanisms behind the physical process being studied. For instance, it has been shown that unknown relations between galaxy properties and parameters describing the composition of the Universe can be easily identified by employing machine learning techniques on top of state-of-the-art hydrodynamic simulations \citep{Cosmo1gal}. We believe that machine learning can trigger a revolution in the large variety of areas of cosmology and galaxy formation that deal with high-dimensional data.

An important question in cosmology is: where does most of the information reside?
While for Gaussian density fields (e.g.~our Universe on large scales or at early times) all information can be extracted using the power spectrum, for highly non-Gaussian density fields the optimal estimator is unknown. Many different works have shown that there is a wealth of cosmological information that is located on small scales and that cannot be retrieved by using the power spectrum. This motivates the use of these scales to tighten cosmological constraints. Unfortunately, these scales not only are non-perturbative (i.e.~they require numerical simulations), but they may also be affected by uncertainties in astrophysical phenomena such as supernova and active galactic nuclei (AGN) feedback. Thus, in a conservative scenario one would like to marginalize over these \textit{baryonic} effects.

In an ideal scenario, it would be desirable to constrain both cosmology and galaxy formation parameters with the highest accuracy. This task could be carried out using machine learning methods, that would require being trained using state-of-the-art cosmological hydrodynamic simulations. With those simulations on hand, one could train neural networks to extract the maximum amount of cosmological and astrophysical information from multi-wavelength observations \citep{Multifield}. This would represent the theoretical formalism needed to extract every single bit of information from cosmological surveys.

Achieving this goal requires running large sets of state-of-the-art hydrodynamic simulations and using machine learning to increase the resolution of the simulations. Alternatively, machine learning could be used to accelerate the simulations themselves. Developing and making publicly available standard datasets for cosmological tasks \citep{CMD} will contribute to the development of the machine learning techniques needed to accomplish these tasks.

However, this ambitious approach is subject to some important caveats. First, it heavily relies on the outcome of numerical simulations, that may or may not, overlap with reality in high dimensions or in low-dimensional projections. For instance, can we trust the outcome of state-of-the-art hydrodynamic simulations when considering the space formed by a large number of galaxy properties? Second, a proper quantification of the statistical, and more important, systematics errors \citep{Multifield} are needed to determine the confidence level of potential discoveries associated to this task (e.g. a detection of the sum of the neutrino masses). Addressing these difficult questions may trigger deep developments not only in the field of hydrodynamic numerical simulations (by improving the simulations to match observations), but also the development of machine learning techniques that may be protected against systematics effects such as numerical artifacts and inaccuracies from the simulations. 

With increase in size and complexity of both simulated and observational data, the need for development of more robust machine learning methods, capable of working with multiple datasets at the same time, will increase. This will allow us to combine the knowledge from hydrodynamic simulations and astrophysical surveys, but also learn from the similarities and differences between them. These differences can come from approximations, numerical artifacts, computational constrains, or even unknown physics, not included in the simulations. On the other hand, observational effects, detector errors and even compression and decompression of the  observed data can introduce additional differences. These differences can cause models trained on simulations to suffer from very large decrease in performance, when applied to real data.

It has been shown in Refs.~\cite{CK2021,CK2022} that domain adaptation methods offer great promise for astrophysics and cosmology. Domain adaptation methods~\cite{C2017}, implemented during model training, help extract only features present in all datasets. This helps align data distributions and build a model that works on all datasets at the same time. Future implementation and development of these methods will be important for building more robust machine learning methods that will not be affected by differences between the datasets, or noise and other perturbations. Furthermore, because domain adaptation methods do not require datasets to be labeled, they can be applied on new observational data. This will be crucial for building automated systems initially trained on simulations, but capable of working in real-time during cosmological survey observations.

\section{Survey Operations and Instrument Design}

\subsection{Survey Operations}

Cosmological surveys such as  the LSST will repeatedly  image the sky, returning to the same part of the sky $\sim$1000 times over a period of ten years. 
The cadence of these observations, the depth of each exposure, and the filters through which the data will be taken will all influence the scientific returns of the survey. 
For experiments such as the {\it Vera C. Rubin Observatory}, the need for almost real-time classification of transient events exacerbates the challenges of how to schedule the observations. 
With multiple scientific objectives and observing conditions that  have both short- and long-term coherence timescales  (e.g.\ atmospheric seeing and weather) traditional hand-tuned scheduling  strategies or policies are not feasible. 
Moreover, coordinating observations amongst many telescopes---e.g., for spectroscopic follow-up or multi-wavelength/-messenger measurements---typically requires proposals to time-allocation committees with months of lag time. 
Finally, how do we optimize the balance of execution toward a specific scientific goal and the exploration for new phenomena?

Current methodologies, like those used in the Dark Energy Survey, rely on simulating many years of high-level metrics (e.g., effective seeing and depth) for a single survey run; humans then review statistics of these metrics, update parameters, and re-simulate until a relative optimum can be found.
Adaptive approaches such as feature-based scheduling \cite{2019AJ....157..151N} that are cast as a formal optimization problem \cite{2015arXiv150307170L} in the context of reinforcement learning have shown that they can optimize multiple competing science objectives, and  outperform current  telescope scheduling approaches while retaining the capacity to recover from unscheduled events (e.g.\ instrument failures). 
Unsupervised methods leveraging graph neural networks have been shown in simulations to optimize for scientific objectives without the need for prescribing any observational policy models \cite{Cranmer:2021pve}.
The above references are a few of the rare examples of work that have gone into AI-related algorithmic resource allocation for telescope observations.

Pursuing automated-scheduling is likely to lower costs and increase efficiency of telescope observations.
Consider the time savings for individual-proposal follow-up campaigns, facility queued observations, and even multi-instrument/-site coordination.  
Moreover, it has the potential to improve science by allowing us to get the best data for a given goal: the telescope can focus on fields and objects (of class and quality) that will serve the prescribed scientific goal (including exploration).

There are several key challenges for growing and making the most of automated resource allocation in telescope observations.
For supervised algorithms that are trained on simulated data or previous observations, there will be a bias when applied to the space of unobserved data: domain adaptation is an avenue to pursue here.
Related, interpreting the decisions and policies that are generated by the scheduling algorithms, as well as the downstream economic and logistical consequences, will be essential for verification, human understanding, and safety.
To grow this effort, we envision administering data benchmarks and challenges, performing cost-benefit analyses to understand economics, and scaled testing of algorithms on small and large telescopes, as well as federations of telescopes.

\subsection{Experiment and Instrument Design}

Observational facilities and their instruments are designed to meet scientific requirements. 
Developing the design specifications that match these requirements is a time- and thus human-intensive process.
Due to the evolving demands of scientific discovery---e.g., increased precision and sensitivity of observations---the complexity of instruments is poised to grow significantly; and this could be a significant bottleneck in terms of time and cost for future experiments.

In the traditional design process, somewhat simplistically, humans first generate hypotheses (e.g., dark energy and dark matter affect the expansion of spacetime and thus the distribution of matter), and then they establish quantitative scientific goals (e.g., errors on the dark matter density) to test those hypotheses, generate the summary metrics that come from data sets (e.g., galaxy two-point correlation function), identify the scope and kinds of data that are required to achieve these metrics (e.g., galaxy survey), and then establish the instrument design and observing specifications to achieve these data (e.g., optical telescope and sensor design). 
The requirement flow-down and design processes are not always this linear from step to step: there are often feedback loops that are shorter than the full loop; an experiment typically has multiple competing science goals, and thus metrics.
Each of the components in the requirement flowdown are typically developed independently within different software frameworks---whatever is required for that component. 
Then, inputs and outputs from each are communicated to their neighboring components, and this feedback process continues until design specifications are achieved that meet the scientific requirements. 

Rarely are multiple, let alone all, of the components linked together within a single software framework. 
For example, in many circumstances, people working on one experimental component communicate inputs and outputs via email.
Also, many instrument simulators used for design run primarily or only as standalone programs that require manual operation: ZeMax is used for optics~\cite{ZeMax}, and Comsol for electromagnetic sensors~\cite{Comsol}.
These all present barriers to fast global optimization of instrument and experiment design.

Connecting these components seamlessly within an algorithmic framework has the potential to reduce bottlenecks and permit fast co-optimization of all the experiment and instrument parameters in service of a set of scientific goals. 
We envision the conversion of requirement flowdowns into models and numerical data. 
There are just a few known areas where some of this concept have been implemented.
For example, the optimization of adaptive optics systems has subject of deep learning applications (see \cite{guo2022adaptive} for a recent review).
Many works are exploring the design of optical element configurations with deep neural networks and similarly flexible algorithms ~\cite{cohen22, Cote:19, 10.1117/12.2528866, Hoschel, Yang2017}.
Co-design of instruments and observational strategies are being performed to optimize black hole observations~\cite{2020arXiv200310424S}.
Finally, there is also a burgeoning community of scientists who are approaching design of instruments with machine learning~\cite{MODE:2021yid} or with the outputs of machine learning-pipelines in mind \cite{bundy2019fobos}.

The only full simulation of a cosmic experiment that has so far been developed is the SPectrOscopic Ken Simulation (SPOKES)~\cite{Nord:2016plv}, which was designed to replicate the DESpec prototype experiment: it includes all major components of a spectroscopic cosmic experiment from data acquisition to analysis to dark energy parameter constraints.
Automating hypothesis generation has not yet been implemented in SPOKES, and there has been limited work in this area for physical experiments -- primarily related to the generation of symbolic models ~\cite{2020arXiv200611287C, Udrescueaay2631}. 
When a field of study is faced with a large landscape of possibilities for theoretical development, automating hypothesis generation has the potential to accelerate the search for new ideas to pursue; this may be of significant import for studying dark matter and dark energy.

There are many potential avenues for the continued development of this area.
For example, a project to design a simple experiment (including the instruments) that is optimized toward a scientific goal can provide a conceptual launching-off point: imagine performing the automated design of a telescope system component-by-component, and then doing that all in tandem.
We recommend significant investment in the following: 1) building full digital twins of experiments (including instruments); 2) exploring the capacity of machine learning for optimizing the design of individual instruments and all instruments within an experiment; and 3) developing the capacity for automating hypotheses, which is among the tightest of bottlenecks in scientific discovery.
Also, some cost-benefit analysis should be performed to study the utility and potential economic gains of automating instrument design.

\section{Machine Learning Architectures}

Until recently, machine learning methods applied in cosmology were borrowed wholesale from the computer science literature. The specific needs of the cosmology community combined with the highly structured nature of cosmological datasets necessitates theoretical and methodological developments on the machine learning side specifically tailored for cosmology applications. Here, we outline ongoing as well as necessary efforts in this direction.

The modern deep learning revolution arguably kicked off in 2012 with deep convolutional neural networks like AlexNet showing record-breaking performance on benchmark vision tasks. Since then, convolutional neural networks have found widespread application within cosmology due to the fact that a lot of cosmological data is naturally represented in the image domain. Despite their success, traditional convolutional neural networks can only process rectangular images, restricting their use in the cosmological context. The need to extend desired implicit biases like translation and rotation symmetry to a more diverse set of domains---such as the celestial sphere---has spurred methodological developments in neural network architectures specifically for cosmological applications. Examples of these developments are custom spherical convolutional networks~\cite{2019A&C....27..130P}, efficient translation- and rotation-invariant normalizing flows~\cite{2022arXiv220205282D}, and fixed kernel-based convolutional architectures~\cite{2021arXiv210411244S,2020MNRAS.499.5902C,Valogiannis:2021chp}. The needs addressed by such developments are often distinct from those in mainstream computer vision research, which is predominantly focused on application to natural images.

As outlined above, of particular importance for  cosmology surveys is the development of methodologies for the analysis of time series data (e.g.\ classification of  light curves from Type 1a supernovae). Recurrent Neural Networks \cite{2020ApJ...905...94V},  Long Short-Term Memory Networks \cite{2021arXiv211201541M}, and Transformers \cite{2021arXiv210506178A} have  made significant  advances for classification problems, but sparsely sampled light curves with noisy observations  present many challenges for  standard neural network architectures  common in applications outside of astrophysics. Physics-inspired and physics-constrained neural networks \cite{2021AJ....162..275B}  have shown promise in reducing the dimensionality of the underlying latent space of a network with an associated reduction in the size of data sets needed to train the network. Most of these applications, however, simply adopt existing network architectures without extending them to the specific requirements of cosmology. For example,  many network architectures are sequential and recursive, which does not easily allow parallel computation (to account for the size of cosmological data sets)  or they assume a Markov process, which means that they cannot easily learn long-range dependencies.

A specific effort to bridge methodological developments in the computer science literature with needs within cosmology is therefore necessary. As a concrete example, spherical convolutional neural networks have seen widespread recent adoption for various cosmological applications (e.g., Refs.~\cite{2021PhRvD.104l3526F,2022MLS&T...3aLT03M,2022arXiv220107771F}) due to the existence of easy-to-use implementations tailored to common pixelizations schemes like HEALPix which are widely used in cosmology~\cite{2019A&A...628A.129K, 2019A&C....27..130P}. This showcases the need for continuous integration and cross-talk between the communities.

The reliability of machine learning models for scientific discovery can be interrogated through the development and application of interpretable and explainable ML systems. Although a mature subfield, explainable AI (XAI) has made limited inroads within cosmology~\cite{2021arXiv210709145H,2020arXiv200301926S}. This can at least in part be attributed to the lack of methods that can be robustly applied in typical scenarios of interest; we therefore see a critical need for methodological development in this direction. Post-hoc application of symbolic regression methods to trained neural networks~\cite{2022arXiv220202306L, Wadekar:2020oov, Wadekar:2022cyw, Delgado:2021cuw} towards model interpretability and model discovery has recently demonstrated promise within cosmology, and we expect this direction to continue to flourish in the near future.

\section{Uncertainty Quantification and Bias}

A major historic barrier to using machine learning models such as deep neural networks within cosmology has been the lack of a robust and validatable statistical description of their output. With increasing adoption of these methods within cosmology, and more broadly in the sciences, the need for such descriptions have become acute. Here, we summarize recent developments in uncertainty and bias quantification, and outline necessary developments for the widespread adoption of machine learning models in the cosmological context.

Perhaps the most common task in cosmological data analysis is parameter inference---i.e., describing what a given set of observations tells us about a set of physically-meaningful parameters in a model (for example, the six $\Lambda$CDM parameters). In the supervised learning setting, the simplest and historically most common way to frame this is as a regression task---e.g., teaching a neural network to minimize the mean square error (MSE) between its output and a set of target parameters. With the expressiveness of the neural network and number of simulations under control, it can be shown the minimizing the MSE loss provides the maximum-likelihood estimate (MLE)~\cite{Goodfellow-et-al-2016}. Although statistically well-defined, the MLE and, more generally, any point estimate, is of limited scientific value when the goal is to quantify if, and how well, a given point in the parameter space is compatible with the data. In cosmology, this fact is further exacerbated since physical degeneracies between parameters are ubiquitous, and there is often the need to compare and/or combine results stemming from different experiments and observations.

Several methods for uncertainty quantification have been developed in the machine learning literature, e.g., Monte Carlo dropout~\cite{10.5555/3045390.3045502}, Bayesian neural networks~\cite{10.5555/2986459.2986721}, and deep ensembles~\cite{10.5555/3295222.3295387}. Direct application of these methods within cosmology and their statistical interpretation is often stymied due to a difference in language in how the two fields talk about uncertainty. In particular, \emph{aleatoric}  and \emph{epistemic} uncertainties (sometimes called \emph{data}  and \emph{model} uncertainties, respectively) are often distinguished in the machine learning literature. Although these terms provide useful abstractions, their use in physics can be confusing since they do not neatly map onto the more familiar concept of statistical and systematic uncertainty. 

The difference in perspectives can be at least partially attributed to what kind of uncertainty is most important in either field. In machine learning, \emph{model} usually refers to the trained model (e.g., a neural network), since the data (e.g., natural language or images) is often assumed to be fixed with little insight into its generating process. On the other hand, in physics we often have substantial insight into the data-generating processes, with \emph{model} encompassing the underlying physics processes, including detector effects that led to the generation of the dataset.
In typical machine learning applications, the aforementioned methods are often used to account for the uncertainty in the machine learning model. When adapted for physics applications, a subset can also account for uncertainty associated with the data-generating process. These methods can produce biased or mis-calibrated uncertainty estimates, even when applied to simple physical systems~\cite{Caldeira:2020dsx}.

Probabilistic machine learning methods have recently demonstrated significant promise for scientific applications, and particularly within cosmology. Methods  under the umbrella of \emph{simulation-based inference} (SBI) aim to estimate full Bayesian posteriors, as well as likelihoods or likelihood ratios (and associated confidence regions) in the setting where the likelihood $p(x\mid\theta)$ of the data-generating process associated with data $x$ and parameters $\theta$ is unavailable or intractable, but samples $x,\theta \sim p(x\mid\theta)$ can be simulated from a mechanistic forward model. See Ref.~\cite{Cranmer:2019eaq} for a recent review. These methods are related in lineage to Approximate Bayesian Computation (ABC), which has been extensively used within cosmology in the last few decades. A typical application involves using a domain-specific neural network, or a projected set of summary statistics, as an input to a parametrized classifier or a conditioning context for a density estimator in order to learn an estimator for the quantity of interest (e.g., posterior or likelihood). Indeed, cosmology-inspired research has motivated the development of efficient SBI methods and algorithms~\cite{Alsing:2018eau,2018PhRvD..97h3004C,Alsing:2019dvb,Alsing:2019xrx,DiazRivero:2020oai,Jeffrey:2020itg,Makinen:2021nly}.

Training parametrized classifiers and density estimators is a challenging task in machine learning, and often requires a large number of simulations and/or a post-hoc calibration procedure in order to produce satisfactory results. It was recently demonstrated that typically-used SBI algorithms tend to produce overly confident posterior estimates~\cite{2021arXiv211006581H}---an unacceptable outcome in cosmological applications. It is therefore imperative to perform diagnostic coverage tests (as done in, e.g., Refs.~\cite{Hermans:2020skz,2021arXiv211006581H,Mishra-Sharma:2021nhh}) to ensure that a trained machine learning-based estimator produces well-calibrated and statistically-consistent results, as well as incentivize continued development of simulation-based inference methods towards robust scientific discovery.

\section{Education and Outreach}

Cosmology, and in fact science at large, is at the cusp of a data-driven evolution driven by the ever increasing open-source software and massive public datasets (both real and simulated), as mentioned above. With this revolution, we have the opportunity to create a more equitable scientific community and greatly increase the reach of our scientific impacts (i.e., create an ``inclusion revolution", \cite{2019ASPC..524..199N}). Here we touch upon educational needs for cross-disciplinary scientists and the requirements for effective outreach programs to the broader public.

Cutting-edge advances in data-driven cosmology require increasingly complex techniques. Cosmologists now regularly publish in ML venues such as NeurIPS and ICML (see, e.g., the recurrent Machine Learning and the Physical Sciences Workshop at NeurIPS). Competitive progress therefore requires mastery of physics (theoretical and/or observational), statistics and data science---a tall order for the upcoming generation of scientists. Massive Open Online Courses (MOOCs) and educational-focused journals (e.g., distill.pub) have addressed this need in varied educational backgrounds by providing free, online supplementary courses to a more traditional physics background. However, few resources exist \textit{by} physicists and \textit{for} physicists that detail scientifically-informed, data-driven methodologies most relevant to our field. Interdisciplinary AI centers are increasingly common, as well as open seminar series which are often broadcast online. We encourage these new foci to continue to facilitate open discussion. We additionally encourage opportunities for collaborative development of new curricula for physics-informed ML. Finally, in the aims of diversifying the field, we encourage funding agencies to directly directly track progress on inclusion of funded programs by requiring documentation of diversity, inclusion and equity efforts \cite{2019BAAS...51g..14N}.

Looking beyond the academic community, cosmologists are additionally at the forefront of new opportunities to engage with the broader public, especially those of historically marginalized groups. On the broadest scale, computational capabilities of handheld electronics (e.g., phones, tablets) allows for unprecedented interactivity with high-quality simulations and graphics. 
Cosmologists can also capitalize on the growing the community of online educational tools and communities for ML by providing datasets to these communities. One potential avenue for this is creating public competitions which is free for participants. We highlight two successful competitions: Galaxy Zoo~\cite{Lintott:2008ne} (in which competitors are tasked with classifying the morphologies of galaxies from images) and PLAsTiCC~\cite{LSSTDarkEnergyScience:2020wcc} (in which competitors must classify transients simulated in an LSST-like datastream). Both of these were held on the website Kaggle and garnered hundreds of submissions. However, we do caution that these communities and competitions likely do not reach out to the most marginalized communities, and indeed may contribute to widening the gap between those with and without dedicated educational resources (as is well known for a need for dedicated education and outreach expertise \cite{kizilcec2017closing}).

\section{Workforce Development}

It is undeniable that the rapid progress in AI/ML has been guided by the commercial gains of these techniques, especially in various technology sectors. At the same time, more than 60\% of physics college graduates end up in the private sector; within this fraction, over 50\% of graduates go into engineering or computational sciences \cite{hiscott_2022}. Both academic and industry settings require knowledge of algorithms, high-performance computing, software carpentry, statistics and machine learning. Integrated computational education within physics curricula is increasingly relevant for preparing a tech savvy workforce. While students can supplement their core curricula with computer science courses, focused courses which teach these fundamental skillsets are becoming increasingly necessary \cite{norman2019growing}. 

Industry is also taking advantage of the open-access, large datasets from cosmology and increasingly creating opportunities to collaborate with academics on these problems (e.g., the Pitt-Google LSST Broker being developed for the community). We encourage continued collaboration with industry, through visiting scholar programs or more junior funded research opportunities.

It is increasingly important that a broad and diverse range of people be involved in these new technologies in order to bring critical perspectives to bear.   If innovation is to flourish, the goal must be to nurture critical and questioning perspectives, shaped by a wide breath of experience, particularly experiences different from the traditional norms currently represented in physics, math and data science.  While such a goal is beneficial to the study of these disciplines, it is crucial for the development of applications and technologies that will exploit algorithms developed by any future workforce. As the physics community engages in the education of the next generation to design, develop and build the future tools, a broad perspective must be taken to train and engage students in the ethics and efficacy of these technologies.  Recruiting and retaining currently minoritized groups is an important piece of expanding the innovation quotient of the workforce.

The lack of input from groups who are often minoritized in the fields of physics and data science can lead to algorithms and AI technologies that reflect the biases and prejudice (conscious or unconscious) of a more narrowly experienced community. Algorithms that mimic racial and gender discrimination are rife within our society. Without opportunities to broaden the range of those who design and build these technologies, we will be unable to advance justice in the scientific culture.

\section{Recommendations and Vision}

Machine learning is set to play an increasingly important role in many facets of cosmology in the next decade. We conclude here by outlining a set of recommendations aimed at realizing our vision for the next generation in the development of artificial intelligence applied to cosmology.
We target these recommendations to a broad community of researchers in the high-energy physics and astrophysics communities.
	
\begin{itemize}
\item Incentivize the use of machine learning techniques when there is a specific advantage towards achieving the stated scientific goal over more thoroughly-vetted traditional statistical (e.g., providing the ability to extract more useful information from the data or to emulate expensive simulations).

\item For parameter inference, eschew point estimation in favor of principled uncertainty quantification, e.g. through simulation-based inference techniques. Assess the statistical soundness of inference pipelines through coverage tests ensuring that the derived quantities of interest (e.g., parameter posteriors) are well-calibrated.

\item Encourage the production and public dissemination of general-purpose cosmological simulations geared towards machine learning applications. Enable this through the availability of centralized and decentralized institutional resources (e.g. HPC clusters as well as cloud computing and storage facilities).

\item Incentivize modular and reproducible workflows in cosmological analyses that can be built upon, independently validated, and used for pedagogical purposes.

\item Invest in cross-survey cyberinfrastructure in order to fully exploit synergies between different cosmological surveys for real-time as well as archival multi-messenger analysis.

\item Recognizing that our models for complex processes in the Universe (reflected in, e.g., simulations) are often not perfect, understand and account for systematic differences between the training and the data distributions through, e.g., principled uncertainty propagation, the use of nuisance parameters defining a flexible family of training distributions, or domain adaptation techniques. Develop and use methods that are robust to systematic model mis-specification in order to minimize the risk of biases.

\item Adopt and develop explainable artificial intelligence techniques with the goal of model interpretability and discovery in cosmology.

\item To the extent possible, minimize data augmentation by leveraging machine learning architectures that encode the expected physical symmetries of the problem and the structure of the data domain. Given the diversity of cosmological datasets, develop efficient architectures suited to processing common cosmological data modalities.

\item Nurture, through education and workforce development, career paths for practitioners with expertise overlap in machine learning methods as well as cosmological applications. Develop curriculum-based learning at the intersection of cosmology and data science aimed at PhD-level and early-career researchers.

\item Develop community-driven data benchmarks and challenges with specific scientific goals in mind; these should be fostered by funding agencies to promote participation.

\item Understand and consider our responsibility towards society at large when developing techniques based on artificial intelligence. Critically assess the broader societal (e.g., potential for harm) and environmental (e.g., carbon footprint) impact of research performed at the intersection of artificial intelligence and cosmology.

\end{itemize}

\vspace{0.5cm}
\subsection*{Acknowledgments}
CD is partially supported by NSF grant AST-1813694.
SM is partially supported by the U.S. Department of Energy, Office of Science, Office of High Energy Physics of U.S. Department of Energy under grant Contract Number DE-SC0012567. VAV is partially supported by the NSF grant AST-2108676.
This manuscript has been authored by Fermi Research Alliance, LLC under Contract No. DE-AC02-07CH11359 with the U.S. Department of Energy, Office of Science, Office of High Energy Physics.
This work is supported by the National Science Foundation under Cooperative Agreement PHY-2019786 (The NSF AI Institute for Artificial Intelligence and Fundamental Interactions). 

\bibliographystyle{utphys}
\bibliography{main.bib}

\end{document}